\documentclass[preprint,aps,showpacs,preprintnumbers,amsmath,amssymb,prl]{revtex4}
\usepackage{graphicx}
\usepackage{color}  
\usepackage{dcolumn}
\usepackage{bm}

\begin{document}

\title{Boolean Chaos}

\author{Rui Zhang\footnote{RZ and HLDSC contributed equally to this article}}
\email{rz10@phy.duke.edu}
\author{Hugo L. D. de S. \surname{Cavalcante}$^*$}
\email{hc71@phy.duke.edu}
\author{Zheng Gao}
\author{Daniel J. Gauthier}
\author{Joshua E. S. Socolar}
\affiliation{Duke University, Department of Physics and Center for Nonlinear and Complex Systems, Durham, North Carolina 27708}
\author{Matthew M. Adams}
\author{Daniel P. Lathrop}
\affiliation{Department of Physics, IPST and IREAP, University of Maryland, College Park, Maryland 20742}

\date{\today}

\begin{abstract}
We observe deterministic chaos in a simple network of electronic logic gates that are not regulated by a clocking signal. 
The resulting power spectrum is ultra-wide-band, extending from dc to beyond 2 GHz.
The observed behavior is reproduced qualitatively using an autonomously updating Boolean model with signal propagation times that depend on the recent history of the gates and filtering of pulses of short duration, whose presence is confirmed experimentally. 
Electronic Boolean chaos may find application as an ultra-wide-band source of radio waves. 
\end{abstract}


\maketitle

We show here that a very simple digital electronic device displays a form of deterministic chaos, a dynamical state characterized by a broadband spectrum and rapid divergence of nearby trajectories.  We also show that a modeling framework based on Boolean state transitions with update times determined by signal propagation explains the origin of this novel behavior, which we term ``Boolean chaos.'' Our device may be used as a building block in secure spread-spectrum communication systems \cite{Volkovskii2005}, an inexpensive ultra-wide-band sensor or beacon, or a basis for engineering high-speed random number generators \cite{Reidler2009}.  It can also be used to address fundamental aspects of the behavior of complex networks.

\begin{figure}[!bt]
\includegraphics[scale=1.00]{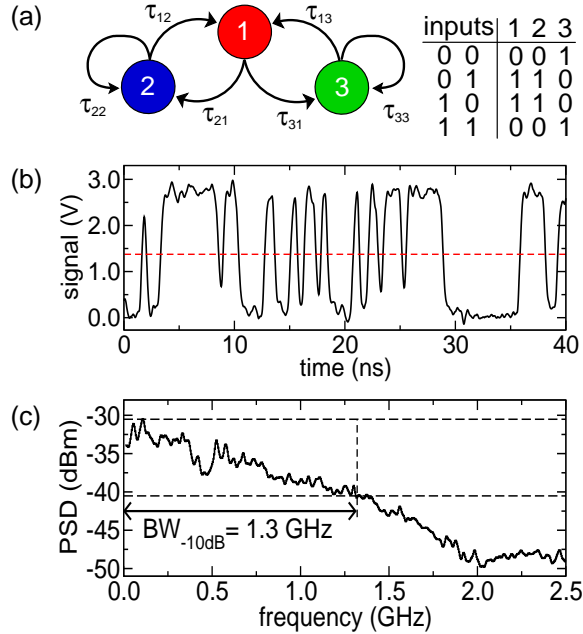}
\caption{\label{fig:network} (Color online) 
(a) Topology of the chaotic Boolean network and truth table for logic operation performed by the nodes 1, 2 ({\sc xor}), and 3 ({\sc xnor}) on their respective inputs.
(b) Temporal evolution and (c) power spectral density (PSD) of the chaotic network for $V_{\textrm{CC}}=2.75$ V with a measurement bandwidth of 1 MHz. }
\end{figure}

Our network consists of three nodes realized with commercially-available, high-speed electronic logic gates. The temporal evolution of the voltage at any given point in the circuit has a non-repeating pattern with clear Boolean-like state transitions, displays exponential sensitivity to initial conditions, and has a broad power spectrum extending from dc to beyond 2 GHz (see Fig.~\ref{fig:network}).  Because the circuit includes feedback loops with incommensurate time delays, it spontaneously evolves to dynamical states with the shortest possible pulse widths, a regime in which time-delay variations generate chaos.  We conjecture that similar behavior will occur in a wide class of systems described by autonomous Boolean networks.

Boolean networks have been studied extensively in a variety of contexts. For systems that display switch-like behavior, such as logic circuits and gene regulatory networks, it is often useful to assume the system variables take only two values ({\it e.\ g.}, ``high'' and ``low'') that are updated according to specified Boolean functions \cite{Jacob1961,Jacob1961a,Davidson2006,Pomerance2009}. Deterministic Boolean models often include an external process such as a clock that synchronizes all the updates or a device that selects a particular order of individual gate updates.  The state space of such models is discrete and finite, and can therefore have only periodic attractors. On the other hand, in many physical or biological systems, information propagates between logic elements with time delays that can be different for each link \cite{Klemm2005,Norrell2007,Glass2005}.  In such systems, the future behavior is determined by specification of the precise times at which transitions occurred in the past, which makes the state space continuous.  The mathematics describing these autonomous Boolean systems is much less developed, though it is known that they can display aperiodic patterns if the logic elements have instantaneous response times \cite{Dee1984,Ghil1985,Ghil2008}.

Ghil and collaborators \cite{Dee1984, Ghil1985, Ghil2008} introduced Boolean delay equations (BDEs) to study Boolean networks of ideal logic elements. They study the dynamics of state transitions, (called events here) in the networks under the hypothesis that the logic gates can process input signals arbitrarily fast.
They consider the behavior to be complex when the event rate per unit time for the whole system grows as a power-law, and predict it can happen for a wide class of Boolean networks.

The complex behavior identified by Ghil leads to an ultraviolet catastrophe that can never occur in an experiment because real logic gates cannot process arbitrarily short pulses.  We find that the predicted complex behavior is replaced by deterministic chaos in our experimental systems and numerical simulations that take into account the non-ideal behaviors described below.  Given the presence of complex behavior in a large class of ideal BDEs, and given our observation of deterministic chaos in a simple experimental example with three nodes, we conjecture that a large class of experimental Boolean networks will display chaos.

The topology of our autonomous Boolean network is shown in Fig.\ \ref{fig:network}(a).  It consists of three nodes that each have two inputs and one output that propagates to two different nodes. 
The time it takes a signal to propagate to node $j$ from node $i$ is denoted by $\tau_{ji}$ ($i,j=1,2,3$).  Nodes 1 and 2 execute the Exclusive-{\sc or} ({\sc xor}) logic operation, while node 3 executes the {\sc xnor} (see truth tables in the Fig.~\ref{fig:network}).  The three-node network has no stable fixed point and always leads to oscillations.  Each time delay comes about from a combination of an intrinsic delay associated with each gate and the signal propagation time along the connecting link, which we augment by incorporating an even number of {\sc not} gates or Schmitt triggers wired in series, either of which act effectively as a time-delay buffer.  We stress that there is no clock in the system; the logic elements process input signals whenever they arrive, to the extent that they are able.

 We observe the dynamics of our network using a high-impedance active probe and an 8-GHz-analog-bandwidth 40-GS/s oscilloscope.  Figure  \ref{fig:network}(b) shows the typical observed behavior when the probe is placed at the output of node 2.  The temporal evolution of the voltage is complex and non-repeating and has clearly defined high and low values, indicating Boolean-like behavior. The rise time of the measured voltage is $\sim$0.2 ns (close to the performance limit of the family of logic gates used in our circuit), and the minimum, typical, and maximum pulse widths in the chaotic time series are 0.2 ns, 2.4 ns, and 12 ns, respectively. In the frequency domain (Fig.\ \ref{fig:network}(c)), the spectrum extends from dc to $\sim$1.3 GHz (10-dB bandwidth).  It is relatively flat up to 400 MHz and decays approximately as the inverse of the frequency from this point on. 

We find that the network dynamics depends on the supply voltage $V_{\textrm{CC}}$ of the logic gates, which we consider as a bifurcation parameter.  Our hypothesis is that the observed dynamics changes with supply voltage because  the different characteristic times of the logic elements, such as the transition time delay, rise and fall times, etc., all depend smoothly on the supply voltage.
To map out a bifurcation diagram for the network, we collect a 1-$\mu$s-long time series of the voltage at node 2 for a fixed value of $V_{\textrm{CC}}$ and transform it into a time series of a Boolean variable $x(t) \in \{0,1\}$ by comparison to a threshold: $x(t) = 0,$ for $V(t) < V_{\textrm{CC}}/2$;  $x(t) = 1,$ for $V(t) \geq V_{\textrm{CC}}/2$ (dashed line in Fig.\ \ref{fig:network}(b)). 
We analyze the resulting Boolean time series to determine the time between successive transitions from low to high values  and plot the observed transition intervals.  We then increase $V_{\textrm{CC}}$ by 5 mV and repeat, starting at $V_{\textrm{CC}}=0.9$ V and ending at 3.3 V. 

\begin{figure}[h,t,b]
\includegraphics[scale=0.90]{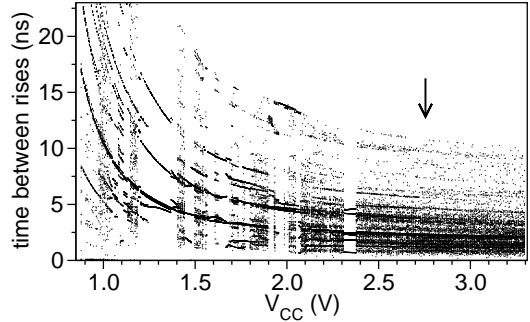}
\caption{\label{fig:bif} Bifurcation diagram of the Boolean network. The arrow indicates the value of $V_{\textrm{CC}}$ giving the  complex behavior shown in Fig.\ \ref{fig:network}(b).}
\end{figure}

As seen in Fig.\ \ref{fig:bif}, the bifurcation diagram shows regions of complex behavior, indicated by a nearly continuous band of points, interspersed by windows of periodic behavior.  The fact that there exist several stable and robust periodic windows demonstrates that our device is not overly sensitive to noise in the voltage.  Furthermore, complex behavior exists over a wide range of supply voltages, especially when $V_{\textrm{CC}}>2.40$ V, where the logic gates are biased to operate at maximum speed.

A signature of chaos is exponential divergence of trajectories with nearly identical initial conditions, which is indicated by a positive Lyapunov exponent. We propose a method to estimate the largest Lyapunov exponent  as follows.
We acquire a long time series of the voltage and transform it to a Boolean variable $x(t)$. Given any two segments of $x(t)$ starting at times $t_a$ and $t_b$, we define a Boolean distance  \cite{Ghil1985} between them by 
\begin{equation}
d(s) = \frac{1}{T} \int_{s}^{s+T}  x(t'+t_a) \oplus x(t'+t_b) dt',
\end{equation}
where $T = 10$ ns is a fixed parameter, $\oplus$ is the {\sc xor} operation, and the Boolean distance $d(s)$ evolves as a function of the time $s$. We then search in $x(t)$ for all the pairs $t_a$ and $t_b$ corresponding to the earliest times in each interval $T$ over which $d(0)<0.01$ ($\ln d(0)<-4.6$). Typically, 3,000 pairs of similar segments are found in a 40-$\mu$s-long time series.
We then compute $\langle \ln d(s)\rangle$, where $\langle\:\: \rangle$ denotes an average over all matching $(t_a,t_b)$ pairs.
\begin{figure}[h,t,b]
\includegraphics[scale=1.0]{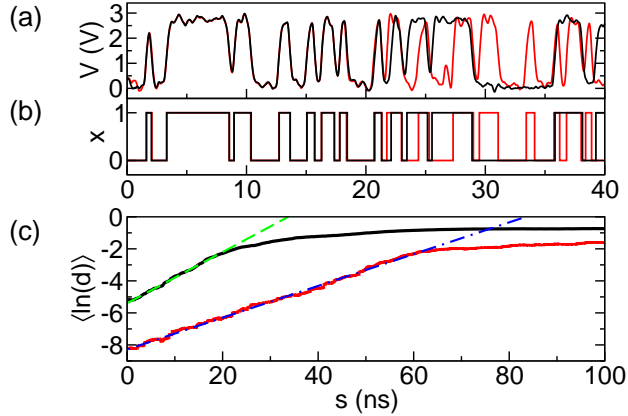}
\caption{\label{fig:Lyapunov} (Color online) (a) Typical segments of similar voltages for $V_{\textrm{CC}}=2.75$ V. (b) The resulting Boolean variables obtained from the voltages in (a). (c) Logarithm of the Boolean distance as a function of time, averaged over the network phase-space attractor for experimental data (black) and simulations (red online).} 
\end{figure}

Figure \ref{fig:Lyapunov}(a) shows two typical segments for the voltages $V(s+t_a)$ and $V(s +t_b)$, and 
Fig.\ \ref{fig:Lyapunov}(b) shows the associated Boolean variables $x(s+t_a)$ and $x(s+t_b)$. 
A visual inspection of the time series on a much finer scale (not shown) reveals that there exist small differences in the timing of events between the two trajectories.  On the scale of the figure, trajectory divergence is noticeable around 20 ns and the two trajectories appear to be uncorrelated after approximately 30 ns. 

To quantify these observations, we determine the largest Lyapunov exponent of the attractor. The solid black curve in Fig.\ \ref{fig:Lyapunov}(c) shows the time evolution of $\langle \ln d \rangle$. It displays an approximately constant slope for times shorter than $\sim$20 ns and, finally, saturates at a maximum value of $\ln 0.5 \approx -0.69$, corresponding to uncorrelated $x(s+T+t_a)$ and $x(s+T+t_b)$. 
To estimate the value of the maximum Lyapunov exponent, we assume that, in the region of constant slope, the divergence of the initially similar segments is exponential, {\it i.e.}, $\ln d(s) = \ln d_0 +\lambda_{ab} s$, where $\lambda_{ab}$ is the local Lyapunov exponent.
The average of $\lambda_{ab}$ over all pairs of similar segments is our estimate of the largest Lyapunov exponent $\lambda$ of the system. 
We find $\lambda= 0.16$ ns$^{-1}$ ($\pm 0.02$ ns$^{-1}$), which demonstrates that the network is chaotic. Our method is based on neighbor searching in the time series of a single element, as described in Ref.\ \cite{Kantz1997}, except that we use the Boolean distance instead of delay-coordinates.

To test our analysis method, we set $V_{\textrm{CC}}$ to place the system in a nearby periodic window (2.35 V) and repeat our analysis.  We find that the Boolean distance stays small ($\langle \ln d \rangle< -4$), as expected. 
Furthermore, we verify that our signal is not generated by a hypothetical linear amplification of correlated noise by comparison between our experimental data and surrogate data, generated by shuffling the time series while preserving its power spectrum and distribution \cite{Kantz1997}. 

To better understand our observations, we study the Boolean delay equations \cite{Dee1984,Ghil1985} 
\begin{equation}
\begin{array}{ccl}
 x_1(t) & = & x_2(t-\tau_{12}) \oplus x_3(t-\tau_{13}), \\
 x_2(t) & = & x_1(t-\tau_{21}) \oplus x_2(t-\tau_{22}), \\
 x_3(t) & = & x_1(t-\tau_{31}) \oplus x_3(t-\tau_{33}) \oplus 1, 
\end{array}
\label{eq:model}
\end{equation}
where $x_i$ is the Boolean state of the $i^{\textrm{th}}$ node 
and the term $\oplus 1$ performs the {\sc not} operation.  The values of $\tau_{ji}$ are given in the last line of Table~\ref{tab:parameters}.  Using initial conditions $\left(x_1(t),x_2(t),x_3(t)\right)=(0,0,0)\ \textrm{for}\ t<0$, we find that the average event rate for $x_1(t)$ (or any of the variables) grows as a power law with exponent $\sim$2, indicating complex network behavior as defined by Ghil \emph{et al.}\ \cite{Ghil1985}. 

This increasingly fast event rate is prevented in the experimental system by the finite response time of the real logic gates.  We find that the dominant contribution to the non-ideal behavior of the network components is due to the series of gates used to generate the delays in the network links; the non-ideal behavior of the {\sc xor} and {\sc xnor} nodes is much smaller and can be modeled by a constant delay after an ideal gate.  To quantify the non-ideal behavior of the links, we measure simultaneously the voltage at the input and output of each link and determine the propagation delay times.  The data display the three non-ideal behaviors:  
(1) short-pulse rejection, also known as pulse filtering, which prevents pulses shorter than a minimum duration from passing through the gate \cite{Klemm2005, Norrell2007}; (2) asymmetry between the logic states, which makes the propagation delay time through the gate depend on whether the transition is a fall or a rise \cite{Norrell2007}; and (3) a degradation effect that leads to a change in the propagation delay time of events when they happen in rapid succession \cite{Norrell2007, Bellido2000}.  We note that these non-ideal behaviors have been proposed for Boolean idealizations of electrical \cite{Bellido2000} and biological networks \cite{Klemm2005, Norrell2007}, suggesting that studies of these effects may have wide application.

\begin{figure}[t,b]
\includegraphics[scale=1.0]{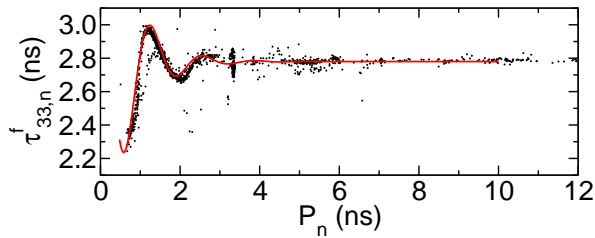}
\caption{\label{fig:degradation} (Color online) Experimentally measured time delay for a transition as it propagates through the delay line 33 (black dots) as a function of $P_n$. Pulses are affected by the degradation effect. The measured values are fit to an empirical expression (solid line) discussed in the text.}
\end{figure}

The non-ideal behaviors are accounted for in our model as follows.  First, following Ref.~\cite{Bellido2000}, we introduce a new variable to describe the degradation effect of a link on signal propagation.  Let $t_n$ be the time that event $n$ occurs at the beginning of a link and let $t'_n$ be the time that the corresponding event is observed at the end of
the link.  Note that $t_n$ does not involve the degradation associated with the link, but $t'_n$ does.  We define
\begin{equation}
P_n \equiv t_n + \tau^k_{ji} - t'_{n-1},
\end{equation}
where $\tau^k_{ji}$ is the nominal time delay on the link for a rising ($k=r$) or falling ($k=f$) event.  Typical behavior of the propagation delay $\tau^f_{33,n} \equiv t'_n-t_n$ for falling events as a function of $P_n$ for link 33 is shown in Fig.\ \ref{fig:degradation}. We fit the experimental data for all links to 
\begin{equation}
\tau_{ji,n}^{k} = \tau_{ji}^{k}+Ae^{-B P_n}\cos(\Omega P_n +\phi)
\label{eq:degradation}
\end{equation}
where $\tau_{ji}^k$, $A$, $B$, $\Omega$, and $\phi$ are fit parameters, and $\tau_{ji,n}$ is the delay of the $n^{th}$ event as it propagates through link $ji$.  The minimum interval $P_{min}$ is determined from the data based on the shortest value for which events are observed.  The only parameter that depends strongly on the specific link and event sign is $\tau_{ji}^k$ (see Table \ref{tab:parameters}).  Based on our fit to the data in Fig.~\ref{fig:degradation} (solid line), we find that the remaining parameters take on values $A=1.28$ ns, $B=1.4$ ns$^{-1}$, $\Omega=4.8$ rad/s, $\phi=0.062$ rad, and $P_{min}=0.48$ ns, which we assume applies to all links in our network.  The next step in our simulation procedure is to solve the ideal Boolean delay equations (\ref{eq:model}) with $\tau_{ji}$ replaced by $\tau_{ji}^k$ as appropriate.  For each event, we evaluate $P_n$.  If $P_n<P_{min}$, both the new event and the previous one are eliminated.  Otherwise, we adjust the newly generated transition time using Eq.~(\ref{eq:degradation}).


\begin{table}
\begin{tabular}{|l|c|c|c|c|c|c|}
 \hline
 \multicolumn{1}{|c|}{$ji$} & \rule{0.2cm}{0cm} 12 \rule{0.2cm}{0cm} &  \rule{0.2cm}{0cm}13 \rule{0.2cm}{0cm} &  \rule{0.2cm}{0cm} 21 \rule{0.2cm}{0cm} & \rule{0.2cm}{0cm} 22  \rule{0.2cm}{0cm} &  \rule{0.2cm}{0cm} 31 \rule{0.2cm}{0cm} & \rule{0.2cm}{0cm} 33 \rule{0.2cm}{0cm} \\
 \hline 
 $\tau_{ji}^r$ (ns) & 3.13 & 4.30 & 3.201 & 2.47 & 3.08 & 3.62 \\
 \hline
 $\tau_{ji}^f$ (ns) & 2.92 & 4.09 & 2.97 & 2.27 & 2.85 & 3.42 \\
 \hline
 \end{tabular}
\caption{\label{tab:parameters} Experimentally measured delay times $\tau^k_{ji}$.}
\end{table}

Using the simulated time series data, we calculate $\langle \ln d \rangle$ (Fig.\ \ref{fig:Lyapunov}(c)) using the initial value of the Boolean distance of $d(0) <0.001$ ($\ln d(0)<-6.9$) for choosing pairs.  We find that $\lambda =  0.10$ ns$^{-1}$ ($\pm 0.02$ ns$^{-1}$), which demonstrates that the model, modified to take into account the non-ideal behaviors of the logic gates, displays deterministic chaos.  Furthermore, the Lyapunov exponents obtained in both the experiment and simulations are very similar, demonstrating that our model captures the essential features of our electronic network.  A systematic study of the effect of each individual non-ideal behavior is beyond the scope of this Letter.

In summary, we observe that an autonomous Boolean network displays deterministic chaos in its sequence of switching times. This behavior is very different from that observed in Boolean networks with periodic or clocked updating, where only periodic behavior is predicted. Our research may have important implications for understanding other networks observed in nature. We note, for example, that chaos was observed in a system of differential equations of a form relevant to the modeling of genetic regulatory networks \cite{Norrell2007}, though the source of chaos was not identified. To make the connection to other natural systems precise, measurements of non-ideal logic elements are needed.  We believe that the three effects identified here are likely to be generic, though they may be difficult to study directly.  Further theoretical study is also needed to determine the extent to which modified Boolean delay equations can serve as a guide for designing and understanding real network behavior.

\begin{acknowledgments}
RZ, HLDSC, ZG, DJG, and DPL gratefully acknowledge the financial support of the Office of Naval Research, grant Nos.\ N00014-07-1-0734 and N00014-08-1-0871, and the advice of John Rodgers. JESS gratefully acknowledges the support of the NSF under grant PHY-0417372.  HLDSC and RZ thank Steve Callender for tips on soldering techniques.
\end{acknowledgments}

\end{document}